\documentclass[aps,prd,tightenlines,showpacs,preprintnumbers,nofootinbib,byrevtex]{revtex4}
\usepackage{amssymb,latexsym}
\usepackage{amsmath,amsbsy}
\usepackage{epsfig,bm}
\DeclareMathOperator{\sign}{\text{sign}}

\DeclareMathOperator{\Tr}{\text{Tr}}

\begin{document}

\def\a{{\alpha}}
\def\b{{\beta}}
\def\d{{\delta}}
\def\D{{\Delta}}
\def\e{{\varepsilon}}
\def\g{{\gamma}}
\def\G{{\Gamma}}
\def\k{{\kappa}}
\def\l{{\lambda}}
\def\L{{\Lambda}}
\def\m{{\mu}}
\def\n{{\nu}}
\def\o{{\omega}}
\def\O{{\Omega}}
\def\S{{\Sigma}}
\def\s{{\sigma}}
\def\th{{\theta}}
\def\x{{\xi}}
\def\Pperp{{\mathbf{P}^{\perp}}}
\def\Pplus{{P^{+}}}
\def\kperp{{\mathbf{k}^{\perp}}}
\def\kplus{{k^{+}}}
\def\dperp{{\mathbf{\Delta}^\perp}}
\def\kpperp{{\mathbf{k}^{\prime \perp}}}
\def\xpp{{x^{\prime \prime}}}
\def\xp{{x^\prime}}
\def\xt{{\tilde{x}}}
\def\eps{{\varepsilon}}
\def\kppperp{{\mathbf{k}^{\prime \prime \perp}}}
\def\Pp{{P^\prime}}
\def\lp{{\lambda^\prime}}

\def\Dslash{\D\hskip-0.65em /}
\def\Dslashe{D\hskip-0.65em /}
\def\Pslash{\ol P\hskip-0.65em /}
\def\lslash{l\hskip-0.35em /}
\def\Pslashe{P\hskip-0.65em /}

\def\ol#1{{\overline{#1}}}

\title{Estimates for Pion-Photon Transition Distributions}
\author{B.~C.~Tiburzi}
\email[]{bctiburz@phy.duke.edu}
\affiliation{Department of Physics\\  
	Duke University\\     
	Box 90305\\
	Durham, NC 27708-0305}
\date{\today}

\begin{abstract}
In the scaling regime, amplitudes for backward virtual Compton scattering 
and hadron annihilation into two photons depend on structure functions
that describe the hadron-photon transition at the partonic level. We construct 
simple analytical models for the vector and axial-vector pion-photon transition distribution 
amplitudes based on double distributions. Via a sum rule, these models are tuned to 
reproduce the pion-photon transition form factors. To obtain reasonable estimates of the
distributions at a low scale, we modify the models to saturate the positivity bounds using 
empirically parametrized parton distributions. We also determine a  model-independent 
contribution to the axial-vector transition distribution using chiral perturbation 
theory.
\end{abstract}


\maketitle

\section{Introduction}

The detailed study of off-diagonal exclusive processes in perturbative QCD%
~\cite{Muller:1994fv,Radyushkin:1996ru,Radyushkin:1996nd,Ji:1997ek,Ji:1997nm}
has lead to a wide variety of theoretical and experimental work.
QCD factorization theorems for such processes enable the physical matrix 
elements to be written as convolutions of perturbatively calculable hard scattering
amplitudes with what are now commonly referred to as generalized parton distributions.  
Typically these distributions arise from hadronic matrix elements that are off-diagonal 
in momentum space and contain quark or gluon operators separated by a light-like distance. 
Parton distributions, which arise in inclusive deep-inelastic scattering, 
involve diagonal matrix elements of light-like separated operators. 
Elastic form factors, on the other hand, are non-diagonal matrix elements of local operators. 
The physics encompassed by generalized distributions is thus that of both inclusive and exclusive processes at the partonic level.
There are a number of insightful reviews on the subject from a variety of perspectives,  
see~\cite{Ji:1998pc,Radyushkin:2000uy,Goeke:2001tz,Belitsky:2001ns,Diehl:2003ny,Belitsky:2005qn}.

In the case of ordinary parton distributions, which we shall generically denote $q(x)$, 
a wealth of phenomenology has resulted from simple analytic parameterizations. For example, 
one might assume at some low input scale $\mu$ the form
\begin{equation}
q(x) = A x^a (1-x)^b
\notag
,\end{equation}
and constrain the parameters $A$, $a$, and $b$ by particle number sum rules and fits to experimental data
at higher scales. Typically there is considerable information buried in these few parameters, since 
one also generally assumes the vanishing of sea quark distributions at 
the input scale, and these distributions are then radiatively generated by perturbative evolution. 
Nonetheless reliable phenomenology has emerged from such Ans\"atze, and global analysis
of structure function data has allowed for better tuned input parameterizations.

For generalized parton distributions, one has functions which we shall generically write as $F(x,\x,t)$. 
In addition to the longitudinal momentum fraction $x$, these functions depend on 
the four-momentum transfer squared, $t$, and the longitudinal momentum transfer, $\x$. 
Lorentz covariance imposes rather stringent constraints on the $x$-moments of $F(x,\x,t)$, 
namely the $n^\text{th}$ moment must be a polynomial in $\x$ of at most $n^\text{th}$ degree. 
The coefficients of these polynomials are in general form factors that depend on $t$. For example, 
the zeroth moment is some form factor $F(t)$. On the other hand, when the four-momentum
transfer goes to zero, we have the reduction relation $F(x,0,0) = q(x)$, which relates
the forward limit of the generalized distribution to the ordinary parton distribution. 
One might be inclined to choose an input parametrization at the scale $\mu$ of the form
\begin{equation}
F(x,\x,t) = q(x) F(t)
\notag
,\end{equation}
and use existing parton phenomenology and form factor data to fix the input distribution. 
Without $\x$ dependence, the polynomiality property of the $x$-moments it trivial. Further
the reduction relation in the forward limit is guaranteed by the form factor's normalization. 
Some problems with the resulting phenomenology are clear: one relies completely on perturbative 
evolution for the distribution to acquire $\x$ dependence, whereas one knows that this dependence
arises at any scale from the interference between light-cone wave-functions of differing
longitudinal momenta~\cite{Diehl:2000xz,Brodsky:2000xy}.  
Furthermore, there is no transparent way to improve the input parameterizations consistent with 
the constraints, i.e. once the parton distribution and form factor are known, the input is fixed.

For generalized parton distribution phenomenology one needs a different way to construct and constrain
input parameterizations beyond the factorized form above. In an alternate approach, one formulates the 
generalized distributions in terms of underlying double distributions~\cite{Radyushkin:1997ki,Radyushkin:1998es}. 
The generalized parton distributions are then obtained as projections of these Lorentz invariant double distributions,
and this feature maintains the polynomiality properties of generalized parton distribution moments.  
Consequently the double distribution formulation is attractive from the perspective of generating input parameterizations. 
Calculations of model double distributions have been 
pursued~\cite{Mukherjee:2002gb,Pobylitsa:2002vw,Tiburzi:2002tq,Tiburzi:2003ja,Tiburzi:2004mh}. 
There is a drawback to using models based on double distributions: the positivity 
bounds~\cite{Radyushkin:1998es,Pire:1998nw,Diehl:2000xz,Pobylitsa:2001nt,Pobylitsa:2002gw,Pobylitsa:2002iu} 
become obscure~\cite{Tiburzi:2002kr}.

In this work we use the formalism of double distributions to provide estimates of pion-photon 
transition distribution amplitudes~\cite{Pire:2004ie}. 
Unlike generalized parton distributions, the off-diagonality of these transition distributions is in particle state, not just momentum. 
But dissimilar to other non-diagonal distributions considered previously~\cite{Frankfurt:1999xe,Pobylitsa:2001cz,Diehl:2003qa}, 
the $t$-channel exchange involved in transition distribution 
amplitudes carries away nearly all the quantum numbers of the initial state.
These distributions enter, e.g.,  in the pion annihilation process, $\pi^+ \pi^- \to \gamma^* \gamma$,  
and backward virtual Compton scattering, $\gamma^* \pi^+ \to \pi^+ \gamma$, 
in the kinematic region where the real photon is nearly collinear to the initial $\pi^+$. 
This region is analogous to the scaling limit of virtual Compton scattering, and QCD factorization applies.
The amplitudes for the annihilation process and backward Compton scattering factorize into a convolution 
of a hard scattering kernel with a soft matrix element, the latter describes the hadron to photon transition~\cite{Pire:2004ie}. 
Analogous proton to pion transition distributions amplitudes appear in the scaling limit of $\ol p N \to \gamma^* \pi$~\cite{Pire:2005ax}.

This paper has the following organization. First in Sec.~\ref{sec:def} we define the 
pion-photon transition distributions of \cite{Pire:2004ie} in terms of double distributions. 
Next in Sec.~\ref{sec:DDcalc}, we present a simple quark model and use it to 
calculate transition double distributions. While the model can reproduce the
pion-photon transition form factors, the partonic content must be modified to give 
reasonable phenomenology for processes at large momentum transfer. In Sec.~\ref{sec:Pos}, 
we investigate the positivity bounds for pion-photon transition distributions. Saturating 
these bounds by including empirically parametrized parton distributions in a factorization 
approximation results in phenomenologically reasonable input distributions satisfying known constraints. 
A conclusion ends the paper (Sec.~\ref{sec:concl}).

\section{Definitions} \label{sec:def}

To begin, we build up the vector and axial-vector pion-photon transition distributions by considering 
matrix elements of towers of twist-two operators. This approach leads us to 
define the transition distribution amplitudes in terms of double distributions.
Consequently Lorentz covariance will be manifest, and this
formulation will ultimately enable us to obtain model estimates that satisfy
known constraints.

\subsection{Vector Operators}
The pion-to-photon matrix elements of vector twist-two operators can be decomposed in a fully Lorentz 
covariant fashion in terms of various twist-two form factors $W_{nk}(t)$, namely
\begin{multline} \label{eqn:moments}
\langle \gamma(P^\prime) | 
\ol \psi (0) \tau^{-} \gamma^{\{\mu}i\tensor D {}^{\mu_1} \cdots i\tensor D {}^{\mu_n\}} \psi(0)
| \pi^{+}(P) \rangle  
 \\ 
= 
- \frac{i e}{f_\pi} \epsilon^*_\nu \ol P_\rho \D_\sigma \varepsilon^{ \nu \rho \sigma \{ \mu } 
\sum_{k=0}^{n} \frac{n!}{ k! (n-k)!} W_{nk}(t) 
\ol P {}^{\mu_1} \cdots \ol P {}^{\mu_{n-k}} 
\left( - \frac{\D}{2}\right)^{\mu_{n-k+1}} \cdots \left( - \frac{\D}{2}\right)^{\mu_{n}\}}
,\end{multline}
where the action of ${}^{\{}\cdots{}^{\}}$ on Lorentz indices produces the symmetric, traceless part of the tensor,
$f_\pi = 132 \, \texttt{MeV}$ is the pion decay constant ,
$\ol P$ is defined to be the average momentum between the initial and final states, 
$\ol P {}^\mu = \frac{1}{2} ( P' + P)^\mu$,
and $\D$ is the momentum transfer, $\D^\mu = ( P' - P)^\mu$, with $t = \D^2$.  
The field $\psi$ is an iso-doublet of the up and down quark fields, and $\tau^\pm$ are the usual
isospin raising and lowering operators.  
$T$-invariance restricts the form factors to be real, but places no further restriction on the index $k$ in the sums.

Ordinarily ambiguity is present in such decompositions of twist-two matrix elements~%
\cite{Polyakov:1999gs,Belitsky:2000vk,Teryaev:2001qm,Tiburzi:2004qr}. 
This situation stems from the generality of writing down form factors of currents 
that are not conserved, and leads to the freedom to define various distributions with the same physical content.  
Here, however, there is only one possible grouping of Lorentz structures in the above decomposition 
due to parity. Nonetheless the twist-two currents are not conserved, 
and there is no ambiguity in the above decomposition into form factors $W_{nk}(t)$.\footnote{%
Electromagnetic gauge invariance is maintained for each twist-two operator as can be 
verified by replacing $\epsilon^*_\nu$ with $P'_\nu$.} 
We now define a double distribution $W(\b,\a;t)$ that is the generating function of the twist-two form factors
\begin{equation} \label{eqn:generate}
W_{nk}(t) = \int_{-1}^{1} d\b \int_{-1 + |\b|}^{1 - |\b|} d\a \, 
\b^{n - k} \a^k W(\b,\a;t)
.\end{equation} 
There is no $\a$-symmetry of this double distribution because the values of $k$ are unrestricted in 
Eq.~\eqref{eqn:moments}.

With the operator product expansion, we can relate the moments in Eq.~\eqref{eqn:moments} to matrix elements of
bi-local operators. By construction, the double distributions appear in the decomposition of the light-like separated 
quark bilinear operator
\begin{equation} \label{eqn:bilocal}
\langle \gamma(P^\prime) | 
\ol \psi \left( - z/2 \right) 
\tau^- \rlap \slash z
\psi \left( z/2 \right) 
| \pi^+(P) \rangle 
=
\frac{i e}{f_\pi} z_\mu \epsilon^*_\nu \ol P_\rho \D_\sigma \varepsilon^{\mu \nu \rho \sigma}
\int_{-1}^{1} d\b \int_{-1 + |\b|}^{1 - |\b|} d\a \;
e^{ - i \b \ol P \cdot z + i \a \D \cdot z / 2}
W(\b,\a;t)
,\end{equation}
where $z^\mu$ is a light-like vector.

The light-cone correlation function is obtained by Fourier transforming with respect to the light-cone separation $z^-$
\begin{equation} \label{eqn:lcc}
\mathcal{M}_V(x,\x,t) = \frac{1}{2} \int \frac{dz^-}{2\pi} 
e^{i x \ol P {}^+ z^-}
\langle \gamma(P^\prime) | 
\ol \psi \left( - z^-/2 \right) \tau^- \gamma^+ \psi \left( z^-/2 \right) 
| \pi^+(P) \rangle
.\end{equation}
Above, the variable $\x$ is the skewness parameter defined by $\x = - \D^+ / (2 \ol P {}^+)$. 
Unlike the case of generalized parton distributions, we cannot assume without loss of generality that $\x > 0$
because time-reversal does not relate the negative $\x$ distributions to those at positive $\x$. 
The correlation function in Eq.~\eqref{eqn:lcc} can be written in terms of the vector pion-photon
transition distribution amplitude $V(x,\x,t)$~\cite{Pire:2004ie}
\begin{equation} \label{eqn:lccor}
\mathcal{M}_V(x,\x,t) = 
\frac{i e}{2 \ol P {}^+ f_\pi}   
\epsilon_\nu^* \ol P_\rho \D_\sigma \varepsilon^{+\nu \rho \sigma}
V(x,\x,t) 
.\end{equation}
Unlike the double distribution, the transition amplitude is a quantity that enters directly 
into convolutions that describe physical processes.
Inserting the decomposition Eq.~\eqref{eqn:bilocal} into the correlator in Eq.~\eqref{eqn:lccor}, we 
can express the transition amplitude as a projection of the double distribution
\begin{equation}
V(x,\x,t) = \int_{-1}^{1} d\b \int_{-1 + |\b|}^{1 - |\b|} d\a \;
\delta(x - \b - \x \a) W(\b,\a;t)
,\end{equation}
from which we interpret the $\x$-dependence of $V(x,\xi,t)$ as arising from different 
slices of the Lorentz invariant double distribution.

The transition distribution satisfies an important sum rule constraint.
As with the case of generalized parton distributions, integrating $\mathcal{M}_V(x,\x,t)$ 
over $x$ produces the local matrix element. Thus we can relate the 
vector transition distribution to the vector form factor $F_V$ accessible in the weak 
decay of the pion. Furthermore matrix elements of the flavor diagonal vector operators
define transition distributions $V^u$ and $V^d$, and these are related to the 
$\pi^0-\gamma$ electromagnetic transition form factor, via~\cite{Pire:2004ie}
\begin{equation} \label{eqn:sumvector}
\int_{-1}^1 dx [Q_u V^u(x,\x,t) - Q_d V^d(x,\x,t) ] = \sqrt{2} f_\pi F_{\pi\gamma}(t)  
.\end{equation}
In our model, isospin algebra will enable us to utilize this sum rule constraint.

\subsection{Axial-Vector Operators}

The analysis of matrix elements of the axial-vector twist-two operators proceeds quite 
similarly. There are, however, additional contributions from so-called internal 
Bremsstrahlung processes that must be separated out from the matrix elements. 
These contributions can be determined from chiral perturbation theory 
in a model independent way, while the structure dependent terms lead to the axial transition 
distribution which is modeled in subsequent sections.

\begin{figure}
\begin{center}
\epsfig{file=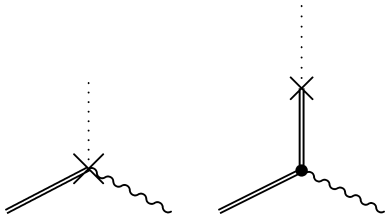,width=2in}
\caption{Tree-level contributions to axial-vector matrix elements in chiral perturbation theory. 
The pion corresponds to the double line, the photon to the wiggly line, while the axial 
current is represented by the dotted line. In the local limit, the cross represents 
the axial current and the depicted contributions are proportional 
to $f_\pi$. These are separated out from the structure dependent axial form factor. 
When the cross represents the axial-vector twist-two operators, the same separation
can be made, but it involves terms  proportional to $f_\pi$ times moments of 
the pion light-cone distribution amplitude.  
}
\label{f:tree}
\end{center}
\end{figure}

To discuss the axial transition distribution, it is useful to recall the form 
of the transition matrix element of the local axial current. For the pion transition 
to a real photon we have, see, e.g.~\cite{Donoghue:1992dd}
\begin{equation} \label{eq:axialcurrent}
\langle \gamma(P^\prime) | 
\ol \psi (0) \tau^{-} \gamma^{\mu} \gamma_5 \psi(0) | \pi^{+}(P) \rangle  
= 
e \, \epsilon^*_\nu 
\left[ 
f_\pi 
\left( 
g^{\mu \nu} - 2 \frac{ \D^\nu \D^\mu}{t - M_\pi^2}   
\right)
+ 
F_A(t) \left( P'^\mu \D^\nu - g^{\mu \nu} P'\cdot \D \right)
\right] 
.\end{equation}
The terms in the above decomposition proportional to the pion decay constant
arise at tree-level in chiral perturbation theory and are depicted in Figure~\ref{f:tree}. 
The last term, usually called the structure dependent piece, involves the 
axial-transition form factor $F_A(t)$. Electromagnetic gauge invariance is 
only present for the structure dependent term. For each process, 
the remaining Bremsstrahlung diagrams are required for gauge invariance. 
The leading contributions in chiral perturbation theory 
stem from the axial-vector operator $A^\mu$
\begin{equation}
A^\mu = 
\frac{f^2}{4} 
\Tr 
( \Sigma^\dagger \tau^- i \partial^\mu \Sigma 
-  
\Sigma \tau^- i \partial^\mu \Sigma^\dagger ) 
\end{equation}
and the electromagnetically gauged axial-vector operator. 
Here $f \approx f_\pi$ at leading order, and $\Sigma$ is an
exponential matrix of the pion fields
\begin{equation}
\Sigma = \exp \sqrt{2} i \, \bm{\pi} \cdot \bm{\tau} / f
.\end{equation}
There are analogous contributions to the axial-vector 
twist-two matrix elements and we shall separate them
out in order to define a structure dependent axial transition 
distribution. To this end, let us determine
the relevant tree-level contribution in chiral perturbation theory.

Following \cite{Arndt:2001ye,Chen:2001eg}, we map the axial-vector twist-two operators
onto those in chiral perturbation theory.  At leading order\footnote{%
At higher orders in chiral perturbation theory, one has contributions with 
two extra derivatives contracted, or equivalently, by the equations of motion, 
insertions of the quark mass matrix. These as well as the one-loop contributions
can be systematically calculated. For meson distribution amplitudes, the one-loop 
corrections have been determined in~\cite{Chen:2003fp}.
We restrict our attention, however, only to those 
terms which survive in the chiral limit.} 
we find
\begin{equation} \label{eq:piontwist}
\mathcal{O}_5^{\mu \mu_1 \ldots \mu_n}  
= 
a^{(n)}_5 \frac{f^2}{4} 
\Tr 
( \Sigma^\dagger \tau^- i \partial^{\{ \mu} i \partial^{\mu_1} \cdots i \partial^{\mu_n\}} \Sigma
- 
 \Sigma \tau^- i \partial^{\{ \mu} i \partial^{\mu_1} \cdots i \partial^{\mu_n\}} \Sigma^\dagger  ) 
,\end{equation}
where the $a_5^{(n)}$ are a set of low-energy constants which in general are not constrained by 
chiral symmetry. They are moments of the pion light-cone distribution amplitude $\phi_\pi(x)$~\cite{Efremov:1979qk,Lepage:1980fj}, 
specifically
\begin{equation} \label{eq:pionmoments}
a_5^{(n)} = \int_0^1 dx \left( x - \frac{1}{2} \right)^n \phi_\pi(x)
.\end{equation} 
We demonstrate this as follows~\cite{Chen:2003fp}. Using the operators in Eq.~\eqref{eq:piontwist}, we determine
\begin{equation}
\langle 0 | 
\ol \psi (0) \tau^{-} \gamma^{\{\mu} \gamma_5 i\tensor D {}^{\mu_1} \cdots i\tensor D {}^{\mu_n\}} \psi(0)
| \pi^{+}(P) \rangle 
= i \, f_\pi \, a_5^{(n)} P^{\{ \mu} P^{\mu_1} \cdots P^{\mu_n \}}
.\end{equation}
Contracting with the light-like vectors $z_\mu z_{\mu_1} \cdots z_{\mu_n}$, and utilizing the operator 
product expansion, we arrive at
\begin{equation}
\langle 0 | \ol \psi \left( - z/2 \right) \tau^- \rlap \slash z  \gamma_5 \psi \left( z/2 \right) | \pi^+(P) \rangle
= i f_\pi \, P \cdot z \int_0^1 d\b \,  e^{- i \left(\b - \frac{1}{2} \right) P \cdot z} \phi_\pi (\b)
.\end{equation}
And thus the familiar definition~\cite{Muller:1994cn} of the pion distribution amplitude emerges via a Fourier transformation,  namely
\begin{equation}
\int \frac{d z^-}{2 \pi} e^{i \left(x - \frac{1}{2}\right) z^- P^+ }
\langle 0 | \ol \psi \left( - z^-/2 \right) \tau^- \rlap \slash z  \gamma_5 \psi \left( z^-/2 \right) | \pi^+(P) \rangle
= i f_\pi \, \phi_\pi(x).
\end{equation}
Integration over $x$ produces the matrix element of the axial current which defines the pion decay constant. 
Thus we require the normalization
\begin{equation}
\int_0^1 dx \, \phi_\pi (x) = 1
,\end{equation}
which tells us further that $a_5^{(0)} = 1$, see Eq.~\eqref{eq:pionmoments}.

Armed with the leading-order operators $\mathcal{O}_5^{\mu \mu_1 \ldots \mu_n}$ 
in Eq.~\eqref{eq:piontwist}, we can calculate the tree-level
chiral contributions to the axial twist-two matrix elements shown in Figure~\ref{f:tree}. 
The first diagram, for example, can be determined by electromagnetically gauging the operators 
in Eq.~\eqref{eq:piontwist}, and then calculating the axial pion-to-photon transition matrix element.
One finds
\begin{equation}
\langle \gamma(P^\prime) | \mathcal{O}_5^{\mu \mu_1 \ldots \mu_n} | \pi^{+}(P) \rangle 
= e\, f \, (n+1) a_5^{(n)} \epsilon^{* \{ \mu} P^{\mu_1} \cdots P^{\mu_n\}}
.\end{equation}
The factor of $(n+1)$ tells us that ultimately this set of diagrams will be related 
to the derivative of the pion distribution amplitude.  The leading-twist contribution from such terms, 
however, is a gauge artifact because $z \cdot \epsilon^* = 0$ in light-cone gauge. 
Thus in the decomposition of axial twist-two transition matrix elements,  
we should ignore structures proportional to $\epsilon^{* \mu}$. The contribution at leading twist from the 
second diagram is not a gauge artifact, and can be calculated using 
the pion electromagnetic current and the operators $\mathcal{O}_5^{\mu \mu_1 \ldots \mu_n}$
in Eq.~\eqref{eq:piontwist}. This calculation enables us to write down the 
decomposition of axial twist-two matrix elements that is the analogue of Eq.~\eqref{eq:axialcurrent}: 
\begin{multline} \label{eqn:moments2}
\langle \gamma(P^\prime) | 
\ol \psi (0) \tau^{-} \gamma^{\{\mu} \gamma_5 i\tensor D {}^{\mu_1} \cdots i\tensor D {}^{\mu_n\}} \psi(0)
| \pi^{+}(P) \rangle  
 \\ 
=  
e\, \epsilon^*_\nu \Bigg[
- 2 f_\pi \, a_5^{(n)} \frac{\D^\nu}{t - M_\pi^2} \D^{\{ \mu} (-\D)^{\mu_1} \cdots (-\D)^{\mu_n \}}
\\
+  \frac{1}{f_\pi}
\D^\nu P'^{\{\mu}
\sum_{k=0}^{n} \frac{n!}{ k! (n-k)!} B_{nk}(t) \, 
P' {}^{\mu_1} \cdots  P' {}^{\mu_{n-k}} 
\left( - \frac{\D}{2}\right)^{\mu_{n-k+1}} \cdots \left( - \frac{\D}{2}\right)^{\mu_{n}\}}
\Bigg]
,\end{multline}
where we have omitted any terms involving the structure $g^{\mu \nu}$, 
and used $\epsilon \cdot P' = 0$ for the real photon.  Here the $B_{nk}(t)$
are structure dependent form factors of the axial twist-two operators. As with the vector
operators, we define a double distribution $B(\b, \a;t)$ to be the generating function
for these form factors
\begin{equation}
B_{nk}(t) 
= 
\int_{-1}^{1} d\b \int_{-1 + |\b|}^{1 - |\b|} d\a \, \b^{n - k} (\a + \b)^k  B(\b,\a;t)
.\end{equation}
Using this definition and the relation of the $a_5^{(n)}$ to the pion distribution amplitude, 
we find that matrix elements of the bi-locally separated quark operator have the form
\begin{eqnarray}
\langle \gamma(P^\prime) | 
\ol \psi (-z/2) \tau^{-} \rlap \slash z \gamma_5  \psi(z/2)
| \pi^{+}(P) \rangle  
&=& 
- 2 e \, f_\pi \frac{ \D \cdot z \, \D \cdot \epsilon^*}{t - M_\pi^2} 
\int_0^1 d\b \, e^{i \left(\b - \frac{1}{2} \right) \D \cdot z} \phi_\pi(\b)
\notag \\
&& + \frac{e}{f_\pi}
P' \cdot z \, \D \cdot \epsilon^*   \int_{-1}^1 d\b \int_{-1 + |\b|}^{1- |\b|} d\a 
\, e^{ - i \b \ol P \cdot z + i \a \D \cdot z / 2} \, B(\b, \a;t)
.\end{eqnarray}

The light-cone correlation function is obtained by Fourier transforming with respect to the
light-cone separation $z^-$
\begin{equation} \label{eqn:lcc2}
\mathcal{M}_A(x,\x,t) = \frac{1}{2} \int \frac{dz^-}{2 \pi} 
e^{i x \ol P {}^+ z^-}
\langle \gamma(P^\prime) | 
\ol \psi \left( - z^-/2 \right) \tau^- \gamma^+  \gamma_5 \psi \left( z^-/2 \right) 
| \pi^+(P) \rangle
,\end{equation}
and can be written in terms of the pion distribution amplitude and the axial
pion-photon transition distribution $A(x,\x,t)$ as
\begin{equation} \label{eq:axialdistribution}
\mathcal{M}_A(x,\x,t) 
= 
e\, f_\pi \sign (\x) \, \theta( |\x| - |x| ) 
\frac{\D \cdot \epsilon^*}{t - M_\pi^2} \, 
\phi_\pi \left( \frac{x+ \xi}{2 \xi} \right) 
+ 
\frac{e}{2 f_\pi} \D \cdot \epsilon^*  \, ( 1 - \x ) \, A(x,\x,t)
,\end{equation}
where 
\begin{equation}
A(x,\x,t) = \int_{-1}^{1} d\b \int_{-1 + |\b|}^{1 - |\b|} d\a \;
\delta(x - \b - \x \a) B(\b,\a;t)
.\end{equation}
The pion pole term derived above is analogous to the one encountered
in the nucleon's $\tilde{E}(x,\x,t)$ generalized parton distribution~\cite{Frankfurt:1998jq,Mankiewicz:1998kg,Penttinen:1999th}. 
In our case, instead of the pion-nucleon coupling, $g_{\pi N}(0)$, we have the pion electromagnetic form factor, $F(0) = Q_{\pi^+} = 1$.

Integration of the amplitude $\mathcal{M}_A(x,\x,t)$ over the variable $x$ produces the 
plus-component of the local axial current operator. The normalization of the pion distribution 
amplitude guarantees that the pion pole term in Eq.~\eqref{eq:axialcurrent} is correctly reproduced. 
Additionally we arrive at the sum rule constraint that relates the transition distribution of 
Eq.~\eqref{eq:axialdistribution} to the axial transition form factor in Eq.~\eqref{eq:axialcurrent}
\begin{equation}
\int_{-1}^1 dx \, A(x,\x,t) = f_\pi \, F_A(t)
.\end{equation}

\section{Quark Model for the Pion-Photon Transition}\label{sec:DDcalc}

In this section, we use a quark model to calculate pion-photon transition
double distributions. Our main goal is to obtain realistic functional dependence
of these distributions on the momentum-transfer within a covariant framework. 
Moreover these models will serve as the core for constructing phenomenological
estimates of the transition distribution amplitudes by incorporating 
realistic $x$-dependence.

The model consists of a triplet of pions $\bm{\pi}$ and a doublet of constituent quarks $q$. 
Their Lagrangian has the form
\begin{equation} \label{eq:modelL}
\mathcal{L} 
= 
\frac{1}{2} \partial_\mu \bm{\pi} \cdot \partial^\mu \bm{\pi}
- \frac{1}{2} M_\pi^2 \, \bm{\pi} \cdot \bm{\pi} 
+ \ol q \, ( i \rlap \slash \partial - m ) q
- i g \, \bm{\pi} \cdot \ol q \, \gamma_5 \bm{\tau} q
,\end{equation}
where we keep the quarks degenerate and
omit color indices from the quarks, as these will produce only trivial color factors below.
To consider electromagnetic interactions, we must add the photon's coupling to each of the particles 
in the above Lagrangian.
We note that unlike the chirally invariant operators considered in the previous section, the Lagrangian
above does not respect chiral symmetry.\footnote{%
Our choice of a chirally non-symmetric Lagrangian for the constituent quarks is based on practical 
considerations. If one uses chiral quarks~\cite{Manohar:1983md}, the pion couples derivatively and 
the asymptotic fall off of the pion-photon transition form factor is contrary to experiment and perturbative QCD. 
Moreover, the pion-photon transition amplitudes are complicated by divergences in the chiral quark model.
For these reasons we have chosen Eq.~\eqref{eq:modelL} as our model.
} 
In this quark model, the quark-pion coupling $g$ is given by $g = m / f_\pi$. 
This model was used long ago to calculate pion-photon transition form factors~\cite{Cabibbo,Resnick:1975fb,Moreno:1976pe,Moreno:1977kx}.
\begin{figure}
\begin{center}
\epsfig{file=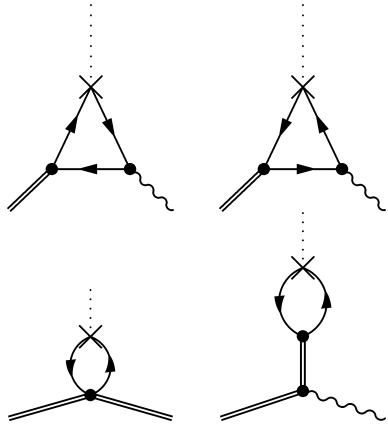,width=2in}
\caption{Impulse approximation to the twist-two matrix elements. Here the twist-two operators with momentum 
insertion are denoted by a cross. 
The initial-state pion has momentum $P$, while the final-state photon has momentum $P'$.
Only the first two diagrams contribute to the vector twist-two matrix elements. The last 
two diagrams depend on the moments of the model's (divergent) distribution amplitude.
These plus the spectator reduced contributions from the first two diagrams must be 
isolated to deduce the remaining, structure dependent axial transition distribution. }
\label{f:twist}
\end{center}
\end{figure}
Additionally a variety of partonic applications of this quark model have been pursued, 
see e.g., \cite{Frederico:1992ye,Frederico:1994dx}.

To apply the quark model to pion-photon transition distributions we must come up with a 
prescription to handle the QCD matrix elements in Eqs.~\eqref{eqn:moments} and \eqref{eqn:moments2}. 
To make any progress in calculating the double distributions, we must use the parton model simplification for 
the QCD gauge covariant derivative: $D^\mu \to \partial^\mu$.
Further we replace the elementary quark fields $\psi$ with constituent quarks $q$. 
Working in the impulse approximation, we have the quark model contributions depicted in Figure~\ref{f:twist},
and we address the determination of the vector and axial vector distributions separately.

\subsection{Vector Distributions}

For the matrix elements of vector twist-two operators, we only have contributions
from the first two diagrams. Explicitly the first depicted diagram has the form
\begin{equation} \label{eqn:impulse}
- i e g \, N_c \, Q_d \, \epsilon^*_\a \, \int  \frac{d^4 k}{(2 \pi)^4} 
\frac{ \Tr[(\rlap\slash k + m) \gamma_5 (\rlap\slash k - \Pslashe + m ) \gamma^\a 
(\rlap\slash k + \Dslash + m)
\Gamma^{\mu \mu_1 \ldots \mu_n}]}
{[k^2 - m^2 + i \varepsilon] \, [(k+\D)^2 - m^2 + i \varepsilon] \, [(k - P )^2 - m^2 + i \varepsilon]}
,\end{equation}
where $N_c$ is the number of colors. 
Above we have already performed the trace over isospin and color indices.
The symmetric, traceless tensor $\Gamma^{\mu \mu_1 \ldots \mu_n}$ 
in Eq.~\eqref{eqn:impulse} is the momentum-space transcription of the parton model operator
$\mathcal{O}^{\mu \mu_1 \ldots \mu_n}$ and is given by 
\begin{equation} \label{eqn:GAMMA}
\Gamma^{\mu \mu_1 \ldots \mu_n} = \gamma^{\{ \mu} (k + \D/2)^{\mu_1} \cdots (k + \D/2)^{\mu_n \}}
.\end{equation}

Deriving the double distribution can be done directly by casting the above contribution into
the form of Eq.~\eqref{eqn:moments}, and then using the definition in Eq.~\eqref{eqn:generate}.
Taking the contribution from both the first and second diagrams, we find the vector transition
double distribution has the form
\begin{equation} \label{eqn:vector}
W(\b,\a;t) 
= 
\frac{m^2}{4 \pi^2} 
\Bigg\{ 
m^2 + \frac{1}{2} \b ( \a + \b - 1) M_\pi^2  - [1 - \a^2 - \b ( 2 - \b)] \frac{t}{4}
\Bigg\}^{-1}
.\end{equation}
Notice in the chiral limit $M_\pi \to 0$, the invariant masses of the initial and final states
are the same and consequently $\a$-symmetry of the double distribution emerges. 
Using isospin algebra, we can calculate the distributions for flavor diagonal twist-two operators. 
We find
\begin{equation}
W^u(\b,\a;t) = W^d(\b,\a;t) = 2 W(\b,\a;t), 
\end{equation}
where $W(\b,\a;t)$ is the flavor non-diagonal distribution given in Eq.~\eqref{eqn:vector}.
Using these relations we can make contact with the sum rule in Eq.~\eqref{eqn:sumvector}, namely
\begin{equation}
F_{\pi\gamma}(t) = \frac{\sqrt{2}}{f_\pi} \int_0^1 d\b \int_{-1 + \b}^{1 - \b} d\a \, W(\b, \a; t) 
.\end{equation}
In the limit $M_\pi \to 0$, we find that the quark model reproduces the normalization of the 
form factor $F_{\pi\gamma}(0) = (2 \sqrt{2} \pi^2 f_\pi)^{-1}$ required by the axial anomaly in the
chiral limit~\cite{Adler:1969gk,Bell:1969ts,Brodsky:1981rp}. 
Notice the variable $\b$ is strictly positive because there are no explicit sea quark contributions
in this model.

\begin{figure}
\begin{center}
\epsfig{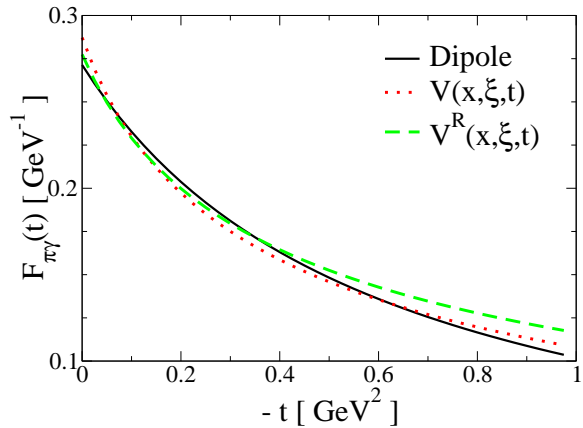}
\caption{Comparison of form factor fits. We compare fits of model pion-photon transition form 
factors to the empirically parametrized form in Eq.~\eqref{eq:empfit}. Fits for the basic model 
$V(x,\x,t)$ derived from Eq.~\eqref{eqn:vector} and the more realistic model $V^R(x,\x,t)$ 
in Eq.~\eqref{eqn:factor} are shown.
}
\label{f:formfit}
\end{center}
\end{figure}

At this stage, we shall fit the constituent quark mass parameter $m$ by requiring that the 
model pion-photon transition form factor comes close to the experimentally parametrized form.
In~\cite{Gronberg:1997fj}, a simple dipole form is used to describe the experimental data
\begin{equation} \label{eq:empfit}
F_{\pi \gamma}(t) = \frac{F_{\pi \gamma}(0)}{1 - t / \Lambda^2}
,\end{equation}
where the dipole mass is $\Lambda = 776 \, \texttt{MeV}$. We will tune the quark mass
at this stage to reproduce the form factor. Eventually, however, we will augment the model distribution
with empirical parton distributions. This will necessitate that the model parameters be
readjusted.  Using the pion mass $M_\pi = 0.14 \, \texttt{GeV}$, 
we find that the form factor at low momentum transfer is well reproduced for $m = 0.18 \, \texttt{GeV}$. 
The limiting and asymptotic behaviors cannot simultaneously be well described with the model, 
however, our concern is to find viable models only at low momentum transfer and thus we have fit
accordingly. In Figure~\ref{f:formfit}, we show our fit to the experimentally parametrized transition form factor.

\subsection{Axial Distributions}

In considering the matrix elements of the axial-vector twist-two operators, all 
of the diagrams in Figure~\ref{f:twist} contribute. The first two diagrams give
contributions quite similar to that in Eq.~\eqref{eqn:impulse} with the insertion of 
an additional $\gamma_5$ from the twist-two operator. We must be careful to remove
contributions that are not structure dependent, in the sense of the decomposition 
in Eq.~\eqref{eq:axialcurrent}. Contributions from the first two diagrams 
where the spectator's propagator can be reduced (by canceling against factors in the numerator)
are not structure dependent~\cite{Moreno:1976pe}. 
This procedure is complicated by the fact that moments of the pion light-cone distribution amplitude in this model are divergent. 
We implicitly use a regularization scheme to render the distribution amplitude finite, and remind the 
reader of this by showing the third diagram in the Figure. The exact form of the scheme used is irrelevant;
subsequent regularized contributions from the fourth diagram, and those reduced contributions from the
first two diagrams are completely discarded because they are not structure dependent. In essence, we 
do not care what the model says about the pion light-cone distribution amplitude, for we have 
already determined the model independent form for this contribution using chiral perturbation theory, 
cf. Eq.~\eqref{eq:axialdistribution}.

These technicalities aside, the structure dependent axial transition distribution amplitude
can be determined from finite parts of the first two diagrams. We find
\begin{equation} \label{eq:axial}
B(\b, \a; t) = 
\frac{3 m^2 ( \a + \b) }{4 \pi^2} 
\Bigg\{ 
m^2 + \frac{1}{2} \b ( \a + \b - 1) M_\pi^2  - [1 - \a^2 - \b ( 2 - \b)] \frac{t}{4}
\Bigg\}^{-1}
,\end{equation}
which is very similar to the vector transition double distribution. The differences result
from the differing spin structure factors present in the interference of the valence light-cone 
wave-functions of the pion and photon, and an overall numerical factor that depends on how the 
charge factors from the first two diagrams combine.
For the parameter value $m = 0.18 \, \texttt{GeV}$, we find $F_A(0)/ F_V(0) = 0.98$
and $F_A(0) = 0.026$. These are consistent with earlier findings for this model~\cite{Moreno:1976pe},
but are more than a factor of two too large compared with experiment~\cite{Frlez:2003pe}.

At this point, we do not plot the vector or axial-vector transition distribution amplitudes 
derived in this model. The main goal of this section has been to determine the $t$-dependence
of the vector and axial form factors in a covariant manner.
We will obtain realistic estimates of transition distributions based on these models and  further 
considerations below.

\section{Positivity and Input Parameterizations} \label{sec:Pos}

The vector and axial vector double distributions determined above are presumably at some very low hadronic scale intrinsic to the model. 
It is not clear how to determine this scale from the model transition distributions alone. One could
in turn calculate the pion parton distributions in the model, and then use the evolution equations 
to evolve empirical parameterizations down to a scale where the first few moments of the model pion parton 
distribution agree and identify this scale with the model. This procedure is not unique; many models can reproduce
the empirical quark distributions at higher scales.\footnote{%
There is no guarantee, moreover, that the scale of the implicit quark-distribution of the photon
is not mismatched with the determined scale of the pion's quark distribution. One should thus calculate
the model's quark distributions for both the pion and photon and determine their mutual scale
by a simultaneous fit.
}
Also the use of perturbative evolution is questionable at best at low scales. 
While the evolution kernels for generalized parton distributions are known at next-to-leading order~%
\cite{Belitsky:1998gc,Belitsky:1998uk,Belitsky:1999gu,Belitsky:1999fu,Belitsky:1999hf}, 
we will not use the evolution approach for our simple model. 
Perturbative evolution cannot generate the non-perturbative small-$x$ physics which our model lacks,
and we believe the small-$x$ physics is crucial for relating to data in the scaling regime. 
The relevant leading-twist amplitudes are convolutions of transition distribution amplitudes 
weighted by a hard scattering kernel that emphasizes a region where either the initial- or final-state wave-function 
is evaluated near the end-point. In fact, the imaginary part of the amplitude
is directly proportional to an overlap of light-cone wave functions, where either the initial or final state has $x=0$.

In order to be useful for phenomenological estimates, a more pragmatic 
solution amounts to augmenting the model with realistic parameterizations of the pion and photon quark distributions
at a low scale.
We shall do this by implanting the realistic distribution via factorization of the $\b$-dependence of the double
distributions~\cite{Mukherjee:2002gb}. 
This choice, while unrealistic for the double distributions, may indeed 
be less problematic for the transition distribution amplitudes since there is ample allowance for interplay of 
$x$, $\x$, and $t$ dependence. From the perspective of model building, we are merely attempting to satisfy 
the known constraints and generate an input distribution at a moderately low scale.
One can then evolve upward to the scale relevant to make contact with data and eventually
learn how to improve the input parametrization.

To implant realistic partonic behavior and estimate the transition distribution amplitudes, 
we shall use a factorization Ansatz motivated by the positvity constraints.
Just like generalized parton distributions, the transition distribution amplitudes
satisfy constraints arising from the norm on Hilbert space: the positivity bounds 
\cite{Radyushkin:1998es,Pire:1998nw,Diehl:2000xz,Pobylitsa:2001nt,Pobylitsa:2002gw,Pobylitsa:2002iu}. 
Using the method described in \cite{Pire:1998nw}, we
can derive the basic bounds satisfied by both the vector and axial vector amplitudes. 
For the positively charged pion, we find
\begin{equation} \label{eqn:positivity}
\theta ( x - \x) \Big| \mathcal{M}_{V,A}(x,\x,t) \Big| \leq 
\sqrt{ d^\gamma\left( \frac{x- \x}{1 - \x}\right) u^{\pi^+}\left( \frac{x + \x}{1 + \x}\right)}
.\end{equation}
Here we have used $f^H(x)$ to denote the parton distribution of the flavor $f$ in the state $H$.
The vector and axial-vector distributions derived in the quark model above
self-consistently satisfy the positivity bounds. That is, the constraints in Eq.~\eqref{eqn:positivity}
are met using the quark distributions of the pion and photon determined within the model.\footnote{%
This can be easily understood due to the simplicity of the quark model used.
When one confronts the light-cone correlation functions $\mathcal{M}_{V,A}$ directly 
instead of via the double distributions, one finds that the contributions for $x>\x$ stem 
from the light-cone time-ordering where the spectator quark is on shell. The remaining 
two propagators then reduce to the model light-cone wave-functions for the initial and final states.
A simple application of the Cauchy-Schwarz inequality gives the bound in Eq.~\eqref{eqn:positivity},
because the quark distributions are transverse momentum integrals of the squares of light-cone
wave-functions.}  
For useful phenomenology, however, these model parton distributions must be modified,
and hopefully in such a way that the bound in Eq.~\eqref{eqn:positivity} remains satisfied.
As an inspired guess, we choose to saturate the bounds in the forward limit 
by making a factorization Ansatz of the form
\begin{equation} \label{eqn:factored}
W^R(\b,\a;t) 
= 
N_V \, \frac{W(\b,\a;t)}{\int_{-1 + \b}^{1 - \b} d\a \, W(\b,\a;0)} \sqrt{d^\gamma\left( \b \right) u^{\pi^+}\left( \b \right)}
,\end{equation}
for the vector distribution and
\begin{equation} \label{eqn:factored2}
B^R(\b,\a;t) 
= 
N_A \, \frac{B(\b,\a;t)}{\int_{-1 + \b}^{1 - \b} d\a \, B(\b,\a;0)} \sqrt{d^\gamma \left( \b \right) u^{\pi^+} \left( \b \right)}
,\end{equation}
for the axial-vector distribution. Here $N_V$ and $N_A$ are constants to be determined below.
When $\{t,\x\} \to 0$, integration over $\b$ and $\a$ of the Ans\"atze yields
\begin{eqnarray} 
V^R(x,0,0) &=&  N_V \sqrt{ d^\gamma \left( x \right) u^{\pi^+} \left( x \right)},  \notag \\
A^R(x,0,0) &=&  N_A \sqrt{ d^\gamma \left( x \right) u^{\pi^+} \left( x \right)} \label{eqn:result}
,\end{eqnarray}
where $V^R(x,\x,t)$ and $A^R(x,0,0)$ are the realistic vector and axial-vector transition distribution 
amplitudes formed from the factorized double distributions. 
From Eq.~\eqref{eqn:result}, one sees our attempt to input realistic
light-cone wave-functions for the pion and photon consistent with positivity. 
The resulting form is reasonable provided there is little interference arising from the 
transverse momentum components of the wave-functions. 
We are not advocating that this is the case. We are only estimating the distributions to the best of 
our ability given reasonable $x$-dependence of parton distributions. Improving these 
estimates sensibly is difficult and the subject of an ongoing investigation. 
In fact for our model, the $\a$ integration in Eq.~\eqref{eqn:factored2} introduces 
a singularity proportional to $1/\b$ in the double distribution. This suggests that 
interference between the various Fock components in the axial distribution cannot be neglected. 
In our model for the axial distribution, saturation of the positivity bounds in the forward limit is not possible. To tame
the singularity, we shall add a factor of $\b$ to the numerator of Eq.~\eqref{eqn:factored2} 
as the simplest way to introduce the requisite interference: 
\begin{equation} \label{eqn:factoredfix}
B^R(\b,\a;t) 
= 
N_A \, \frac{\b \, B(\b,\a;t)}{\int_{-1 + \b}^{1 - \b} d\a \, B(\b,\a;0)} 
\sqrt{d^\gamma \left( \b \right) u^{\pi^+} \left( \b \right)}
.\end{equation}

In the literature, there exists a number of analytic parameterizations of quark distributions 
that have been constrained to fit data. Our preference is to choose those models which are 
at a reasonably low scale, so that evolving up to the relevant scale for comparison with experiment 
allows some of the crudeness in the factorization Ans\"atze to be evolved away.
For the pion, in Ref.~\cite{Gluck:1999xe} simple analytic parameterizations of the valence and sea quark 
distributions are presented. As the sea quark distributions are unconstrained experimentally, they have
been related, using a constituent quark model, to the radiatively generated parton distributions in 
the proton, see~\cite{Altarelli:1995mu,Gluck:1997ww}. 
For our model estimates, we need the up distribution in the positively charged pion,
\begin{equation}
u^{\pi^+}(x)  = \frac{1}{2} v^{\pi^+}(x) + \ol q^{\pi^+}(x)
.\end{equation}
This distribution is given in terms of the valence and sea distributions defined in \cite{Gluck:1999xe} that are given by
\begin{eqnarray}
v^{\pi^+}(x, \mu^2 = 0.40 \, \texttt{GeV}^2 ) &=& 1.391 \, x^{-0.447} (1- x) ^{0.426}, \\
\ol q^{\pi^+}(x, \mu^2 = 0.40 \, \texttt{GeV}^2 ) &=& 0.417 \, x^{-0.793} ( 1 - 2 .466 \, \sqrt{x} + 3.855 \, x) (1-x)^{4.454}
.\end{eqnarray}
The input scale $\mu^2 = 0.40 \, \texttt{GeV}^2$ is obtained from using next-to-leading order QCD evolution.

For the real photon, we use the analytic parameterization presented in~\cite{Gluck:1999ub}.
The photon's quark distribution consists of a perturbative point-like contribution and a non-perturbative 
hadronic contribution. The former is chosen to vanish at the input scale $\mu^2 = 0.40 \, \texttt{GeV}^2$.
The hadronic piece is determined using a vector meson dominance model~\cite{Schuler:1995fk}, and by relating 
the quark distributions in vector mesons to those in the pion.  For the down quark distribution of the photon
\begin{equation}
d^\gamma (x) = \frac{1}{4 \pi} G_d^2 \, d^{\pi^0} (x)
,\end{equation}
where $G_d^2 \approx 0.25$, and $d^{\pi^0} = \frac{1}{2} ( d^{\pi^+} + d^{\pi^-})$. 
In terms of the parametrized forms for the pion parton distribution used above, 
$d^{\pi^0} = \frac{1}{4} [ v^{\pi^+} + 4 \ol q^{\pi^+}]$.

The terms motivated by the positivity bound in the region $\x < x < 1$ 
all concern the quark distributions of the pion and photon and hence contribute
to the transition distribution amplitudes in the kinematic region $-\x < x < 1$.
Assuming the distributions vanish outside this range is inconsistent with the 
partonic input. 
To arrive at the anti-quark contributions in the kinematic region $-1 < x < \x$, 
we make a similar set of Ans\"atze  by considering the positivity bound
in the region $-1 < x < -\x$. In this region, we have
\begin{equation} \label{eqn:positivity2}
\theta (\x - x) \Big| \mathcal{M}_{V,A}(x,\x,t) \Big| \leq 
\sqrt{ \ol d {}^\gamma\left( \frac{x- \x}{1 - \x}\right) \ol u^{\pi^+}\left( \frac{x + \x}{1 + \x}\right)}
.\end{equation}
In terms of the parameterizations used above, 
$\ol u^{\pi^+}(x) = \ol q^{\pi^+}(x)$ and $\ol d {}^\gamma (x) = d^\gamma (x)$. 
For lack of more information, we are forced to assume the anti-quark contributions have the
same $t$-dependence as the quark contributions, and thus take the anti-quark 
distributions to be of the form
\begin{equation} \label{eqn:factored3}
\ol W {}^R (\b,\a;t)
= 
N_V \, \frac{W(\b,\a;t)}{\int_{-1 + \b}^{1 - \b} d\a \, W(\b,\a;0)} \sqrt{\ol d {}^\gamma\left( \b \right) \ol u^{\pi^+}\left( \b \right)}
,\end{equation}
for the vector distribution and
\begin{equation} \label{eqn:factored4}
\ol B {}^R (\b,\a;t) 
= 
N_A \, \frac{\b \, B(\b,\a;t)}{\int_{-1 + \b}^{1 - \b} d\a \, B(\b,\a;0)} \sqrt{\ol d {}^\gamma \left( \b \right) \ol u^{\pi^+} \left( \b \right)}
,\end{equation}
for the axial-vector distribution.
A factor of $\b$ has been inserted because the axial distribution becomes singular 
if one saturates the positivity bound in the forward limit. 
We have defined these functions for positive $\b$ but have used a barred notation to remind the reader that 
they are anti-quark contributions. The realistic transition distribution amplitudes 
are formed from the barred and unbarred double distributions via
\begin{eqnarray} \label{eqn:factor}
V^R(x,\x,t) 
&=&  
\int_{0}^1 d\b \int_{-1 + \b}^{1 - \b} 
\left[ 
\delta(x - \b - \x \a) W^R(\b,\a;t) - 
\delta(x + \b + \x \a) \ol W {}^R(\b,\a;t)
\right]
,\\  \label{eqn:factor2}
A^R(x,\x,t) 
&=&  
\int_{0}^1 d\b \int_{-1 + \b}^{1 - \b} 
\left[ 
\delta(x - \b - \x \a) B^R(\b,\a;t) - 
\delta(x + \b + \x \a) \ol B {}^R(\b,\a;t)
\right]
,\end{eqnarray}
where the minus sign arises from the opposite ordering of the anti-quark creation and 
annihilation operators, see~\cite{Diehl:2000xz}.

\begin{figure}
\begin{center}
\epsfig{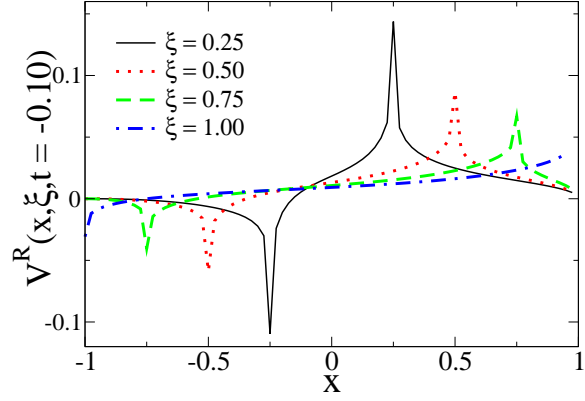}
\caption{Vector pion-photon transition distribution amplitude. 
The function $V^R(x,\x,t)$ is plotted at fixed $t = -0.10 \, \texttt{GeV}^2$
for a few values of $\x$. Notice the distribution is only qualitatively $x$ anti-symmetric. 
}
\label{f:V}
\end{center}
\end{figure}

\begin{figure}
\begin{center}
\epsfig{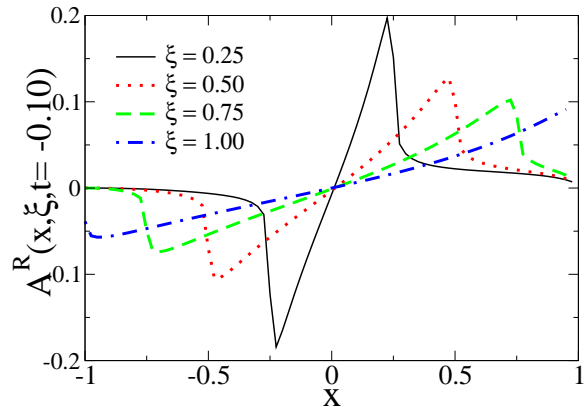}
\caption{Axial-vector pion-photon transition distribution amplitude. 
The function $A^R(x,\x,t)$ is plotted at fixed $t = -0.10 \, \texttt{GeV}^2$
for a few values of $\x$. The distribution is only qualitatively $x$ anti-symmetric 
}
\label{f:A}
\end{center}
\end{figure}

Using these parameterized parton distribution functions 
in the Ans\"atze Eqs.~\eqref{eqn:factored}, \eqref{eqn:factoredfix}, \eqref{eqn:factored3}, and \eqref{eqn:factored4}, 
we can determine the constants
$N_V$ and $N_A$ by using the experimental numbers for $F_V(0)$ derived from
CVC, and $F_A(0)$. Using $F_V(0) = 0.0259$, and $F_A(0) = 0.0115$~\cite{Frlez:2003pe}, we find
$N_V = 0.33$ and $N_A = 0.48$. There is no reason to suppose that the constituent 
quark mass parameter $m$ remains the same when augmenting the double distribution
with parametrized quark distributions. Just as we did for the basic model, we can determine 
$m$ for the realistic model by requiring a reasonable fit to the small momentum transfer 
behavior of the pion-photon transition form factor $F_{\pi \gamma}(t)$. In Figure \ref{f:formfit},
we have shown the form factor determined from $V^R(x,\x,t)$ for the value $m = 0.20 \, \texttt{GeV}$,
as well as the earlier fit, and the empirical dipole form. We are able to match the low momentum 
behavior of the transition form factor. 
Having determined the model parameters for the vector and axial vector distributions in 
Eqs.~\eqref{eqn:factor}, and \eqref{eqn:factor2}, we can now plot the estimates of transition 
distributions amplitudes. The vector distribution $V^R(x,\x,t)$ is plotted as a function of $x$ at fixed $t = -0.10 \, \texttt{GeV}^2$
for a few values of $\x$ in Figure~\ref{f:V}. The same is done in Figure~\ref{f:A} for the function $A^R(x,\x,t)$. 
One should note that in the symmetric kinematics, we have the relation
\begin{equation}
t = - \frac{\bm{\Delta}_\perp^2 - 2 \x ( 1- \x) M_\pi^2}{1 - \x^2}
.\end{equation}
Thus for fixed $t$, we must have
\begin{equation}
- \frac{1 }{1  - 2 M_\pi^2 / t } \leq \x \leq 1
,\end{equation}
and unlike the case of deeply virtual Compton scattering, the maximal
skewness is not kinematically restricted. In the Figures, the minimal 
skewness at $t = -0.10 \, \texttt{GeV}^2$ is $\x_{\text{min}} = - 0.72$. 
We have not plotted the distributions for negative values of $\x$. In our model
the differences between distributions at positive and negative values of $\x$ 
are minimal. The approximate $\x$-symmetry that emerges is due to the factorization
Ans\"atze which evaluate the pion and photon parton distributions symmetrically.

\section{Conclusion} \label{sec:concl}

Above we have written the vector and axial-vector 
pion-photon transition distribution amplitudes in terms of transition double distributions.
These double distributions were defined from the matrix elements of the relevant twist-two operators. 
For the axial-vector transition distribution, we determined the form of a model-independent contribution
depending on the pion light-cone distribution amplitude from chiral perturbation theory.

Using a quark model, we calculated the transition double distributions, and demonstrated that 
the fall-off of the pion-photon electromagnetic transition form factor could be well reproduced 
by tuning the constituent quark mass parameter. 
Motivated by the positivity bounds, we then took a factorized 
Ansatz for the double distributions in order to establish the renormalization scale of the model, 
and to incorporate parameterized parton distributions with realistic $x$-dependence.  
Model parameters were then re-tuned to produce the pion-photon 
electromagnetic transition form factor and the pion axial form factor. 
We then plotted the resultant input distributions which should lead to reasonable 
phenomenological estimates for pion-photon transitions in the scaling regime. 
It would be interesting to compare our estimates with other calculations. There are a 
number of existing realistic studies of pion structure that could be extended to 
transition distributions amplitudes, e.g.~\cite{Petrov:1998kg,Anikin:1999cx,Anikin:2000th,Dalley:2002nj,Dalley:2004rq}.

Improving the factorization Ansatz employed above may ultimately be necessary to best 
describe the $x$, $\x$,  and $t$ dependence of the data. 
There is ample room for improvement, e.g., we assumed the same 
$t$-dependence and absolute normalization for quark and anti-quark contributions.
More realistic models, data from lattice QCD, and experimental data can help to motivate
better input parameterizations. 
Furthermore the only difference between vector and axial vector distributions in our model
arises from the spin structure of the valence light-cone wave-functions, whereas realistically the difference 
stems from the interference from all Fock components. One needs to address whether our approximation is 
sensible for the input distributions at a low scale.
We already found suggestions of sizable interference between Fock components in the axial distribution.
Finally approximate $\x$-symmetry of the distributions emerges due to the simplicity of our Ans\"atze. 
Better input parameterizations must address how to treat the initial and final states differently in a 
covariant fashion.

The study of off-diagonal exclusive reactions leads to phenomenology
involving generalized parton distributions. Unlike ordinary parton distributions, 
input parameterizations cannot simply be written down at will, as constraints on the 
distributions are non-trivial. Using double distributions, we have formulated a basic input 
parameterization for pion-photon transition distribution amplitudes. Contact with future data, 
lattice QCD data, or other models will help one to tune this type of input parameterization. 
This procedure in turn gives one a better sense of how to construct reasonable input
parameterizations of generalized parton distributions in general.

\begin{acknowledgments}
We thank B.~Pire for inciting us to investigate these distributions, and G.~A.~Miller for initial discussions.
This work was funded by the U.~S.~Department of Energy grant DE-FG$02-96$ER$40945$. 
\end{acknowledgments}

\bibliography{lc.bib}

\end{document}